\documentclass[11pt]{article}
\pdfoutput=1
\usepackage{amsmath}
\usepackage{amssymb}
\usepackage{mathtools}
\usepackage{graphicx,bbm,mathrsfs}
\usepackage{nicefrac}
\usepackage{bbm}
\usepackage{geometry}
\usepackage{rotating}
\usepackage{subdepth}
\usepackage{jheppub}

\usepackage{dcolumn}   
\usepackage{bm}        
\usepackage{graphicx,mathrsfs}
\usepackage{nicefrac}
\usepackage{multirow}
\usepackage{xcolor}

\usepackage[nomargin,inline,marginclue,draft]{fixme}
\fxsetup{theme=color}

\def\ie{{\it i.e.}}
\def\eg{{\it e.g.}}
\newcommand{\be}{\begin{equation}}
\newcommand{\ee}{\end{equation}}
\newcommand{\bea}{\begin{eqnarray}}
\newcommand{\eea}{\end{eqnarray}}

\newcommand{\tr}{\operatorname{tr}}

\usepackage{wasysym}

\makeatletter
\g@addto@macro\bfseries{\boldmath}
\makeatother

\title{The Simplified Likelihood Framework}

\author[a]{Andy~Buckley,} 
\author[b]{Matthew~Citron,} 
\author[c,d]{Sylvain~Fichet,}
\author[e]{Sabine Kraml,}
\author[f,g]{Wolfgang~Waltenberger,}
\author[h]{Nicholas~Wardle}

\affiliation[a]{School of Physics \& Astronomy, University of Glasgow, Glasgow, Scotland, UK}
\affiliation[b]{University of California, Santa Barbara, Santa Barbara, California, USA}
\affiliation[c]{Walter Burke Institute for Theoretical Physics,
California Institute of Technology, Pasadena, CA 91125, California, USA}
\affiliation[d]{ICTP-SAIFR \& IFT-UNESP, R.\ Dr.\ Bento Teobaldo Ferraz 271, S\~ao Paulo, Brazil}
\affiliation[e]{Laboratoire de Physique Subatomique et de Cosmologie, Universit\'e Grenoble-Alpes, CNRS/IN2P3, 53 Avenue des Martyrs, F-38026 Grenoble, France}
\affiliation[f]{Institut f\"ur Hochenergiephysik,  \"Osterreichische Akademie der Wissenschaften, Nikolsdorfer Gasse 18, 1050 Wien, Austria}
\affiliation[g]{University of Vienna, Faculty of Physics, Boltzmanngasse 5, 1090 Wien, Austria}
\affiliation[h]{Imperial College London, South Kensington, London, UK}

\emailAdd{andy.buckley@ed.ac.uk}
\emailAdd{matthew.citron@cern.ch}
\emailAdd{sylvain.fichet@gmail.com}
\emailAdd{sabine.kraml@lpsc.in2p3.fr}
\emailAdd{walten@hephy.oeaw.ac.at}
\emailAdd{nckw@cern.ch}

\abstract{We discuss the simplified likelihood framework 
as a systematic approximation scheme for experimental likelihoods such as those originating from LHC experiments.
We develop the simplified likelihood from the Central Limit Theorem keeping the next-to-leading term in the large $N$ expansion to correctly account for asymmetries. Moreover, we present an efficient method to compute the parameters of the simplified likelihood  from Monte Carlo simulations. The approach is validated using a realistic LHC-like analysis, and the limits of the approximation are explored. 
Finally, we discuss how the simplified likelihood data can be conveniently released in the  HepData error source format and automatically 
built from it, making this framework a convenient tool to transmit realistic experimental likelihoods to the community.}

\begin{document} 
\maketitle
\flushbottom


\section{Introduction}

Scientific observations of the real world are by nature imperfect in the sense that they always contain some amount of
uncertainty unrelated to data, the \textit{systematic}   uncertainty. Identifying, measuring and modelling all the sources of systematic uncertainty is an important part of running a scientific experiment. A thorough treatment of such uncertainties is especially important in exploratory fields like particle physics and cosmology. In these fields of research,  
today's experiments  can be of large scale and can contain a huge number of these  uncertainties. In the case of the Large Hadron Collider (LHC) experiments, for instance, the experimental likelihood functions used in Standard Model measurements and searches for new physics can contain several hundreds of systematic uncertainties. 

Although sources of systematic uncertainty can be numerous and of very different nature, a general feature they share is that their most elementary components tend to be independent from each other. This property of independence between the elementary systematic uncertainties has profound consequences, and, as discussed below, is the reason why the approach presented in this work is so effective. Namely, independence of the uncertainties can be used to drastically simplify the experimental likelihood function, for the price of an often-negligible error that will be discussed at length in this paper.

The \textit{simplified likelihood} (SL) framework we present in this paper is a well-defined approximation scheme for experimental likelihoods. It can be used to ease subsequent numerical treatment like the computation of confidence limits,  to allow a uniform statistical treatment of published search-analysis data and to ease the transmission of results between an experiment and the scientific community.
We build on the proposals for approximating likelihoods recently suggested in Refs.~\cite{Fichet:2016gvx,SL_note}, in which promising preliminary results have been shown.

In the context of the LHC, communicating the experimental likelihoods, in their full form or in convenient approximations, was advocated in Refs.~\cite{Kraml:2012sg,Boudjema:2013qla}. 
One possibility is to communicate the full experimental likelihoods via the \texttt{RooFit/Roostats} software framework~\cite{Verkerke:2003ir,Moneta:2010pm}.  
The presentation method we propose in this paper is complementary in that it is technically straightforward to carry out, without relying on any
particular software package. Additionally, the proposal of presenting LHC results decoupled from systematic uncertainties has been pursued in Ref.~\cite{Cranmer:2013hia} in the context of theoretical errors on Higgs cross-sections. For Higgs cross-sections and decays, the combined covariance of the Higgs theoretical uncertainties consistent with the SL framework presented here has been determined in Ref.~\cite{Arbey:2016kqi}.

In this paper we unify and extend the initial proposals of Refs.~\cite{Fichet:2016gvx,SL_note}, and thoroughly test the accuracy of the approximations using simulated LHC searches for new phenomena.
Compared to Refs.~\cite{Fichet:2016gvx,SL_note}, an important  refinement is that we provide a way 
to rigorously include asymmetries in the combined uncertainties, which is useful in order to avoid inconsistencies such as a negative event yield. Technically this is done by taking into account the next-to-leading term in the limit given by an appropriate version of the Central Limit Theorem (CLT).

The paper is organised as follows. 
Section~\ref{se:EL_SL} introduces the formalism and key points of our approach.  
The formal material, including an in-depth discussion of the next-to-leading term of the CLT and the derivation of the SL formula, is presented in Section~\ref{se:SL_theory}. 
Practical considerations regarding the SL flexibility and the release of the SL via HepData are given in 
Section~\ref{se:practice}.
Finally a validation of the SL framework in a realistic pseudo-search at the LHC is presented in Section~\ref{se:SL_LHC}. 
Section~\ref{se:conclusions} contains our summary and conclusions.
Two appendices give some more useful details: 
Appendix~\ref{app:skew} contains a 1D example of how the skew appears in the asymptotic distribution, and 
Appendix~\ref{app:reference_code} presents a reference implementation of the SL written in Python.

\clearpage
\section{From the experimental likelihood to the simplified likelihood}
\label{se:EL_SL}

This section introduces the formalism and an efficient Monte-Carlo based calculation method. 
Some preliminary remarks are in order.  
From the conceptual point of view, the SL framework relies only on the convergence of the CLT. In practice however, 
the representation of the SL will depend on broad, structural features of the dataset under consideration.  
The case considered in this paper is a set of $P$ independent observables in the presence of $N$ 
independent sources of uncertainties, with $N\geq P$.  
For a dataset with different structure, the SL would take a different form, but this  
is not a fundamental limitation of the approach \textit{per se}. 
Moreover, in the scope of a given problem, \eg\ the search for new physics in our case, additional approximations 
may simplify the formalism. Again, this should not be understood as a fundamental limitation, as such approximations  
could be removed in a different application. A summary of the validity conditions for the SL treated in this paper 
will be given in Section~\ref{se:conclusions}.

In the following, we will focus on the typical experimental likelihood used in searches for new phenomena at particle physics experiments. 
However, as argued above, the SL approach can be easily generalised to other physics contexts.
The data collected in particle physics usually originate from random (quantum) processes, and have thus an intrinsic \textit{statistical} uncertainty--which vanishes in the limit of large data sets. Our interest rather lies in the \textit{systematic} uncertainties, which are independent of the amount of data. 

A likelihood function $L$ is related to the probability Pr to observe the data given a model $\cal M$, specified by some parameters,
\be L({\rm parameters})={\rm Pr}({\rm data}|{\cal M},{\rm parameters})\,.\ee
We denote the observed  quantity as $n^{\rm obs}$ and the expected quantity by $n$, where $n$ depends on the model parameters.   For example, in the case of
a particle physics experiment, these quantities can be the observed and expected number of events that satisfy some selection criteria.
The full set of parameters includes parameters of interest, here collectively denoted by $\bm{\alpha}$, and \textit{elementary} nuisance parameters $\bm{\delta}=(\delta_1,\ldots,\delta_{j}\ldots,\delta_N)^{\rm{T}}$, which model the systematic uncertainties.
In the SL framework, we derive a set of \textit{combined} nuisance parameters $\bm{\theta}$. For $P$ independent measurements, there will be $P$ combined nuisance parameters, $\bm{\theta}=(\theta_1,\ldots,\theta_{I},\ldots,\theta_P)^{\rm{T}}$.

The key result at the basis of  the SL framework is the approximation
\begin{align}
  & L(\bm{\alpha},\bm{\delta} )\pi(\bm{\delta})
   = \prod_{I=1}^P \mathrm{Pr}\Big(  n^{\rm obs}_I\,\Big|\,n_I(\bm{\alpha},\bm{\delta})  \Big) \pi(\bm{\delta}) \label{eq:EL} \\
  & \quad \approx \prod_{I=1}^P \mathrm{Pr}\Big( n^{\rm obs}_I\,\Big|\,a_{I}(\bm \alpha)+b_I(\bm \alpha)\theta_I+c_I(\bm \alpha)\theta_I^2  \Big) \cdot
    \frac{ \mathrm{e}^{{\textstyle-\frac{1}{2}\bm \theta^{\rm T} \bm{\rho}^{-1}(\bm \alpha) \bm \theta}}}{\sqrt{(2\pi)^P }}  \equiv L_{\rm{S}}(\bm{\alpha},\bm{\theta})\,, \label{eq:SL_master}
\end{align}
where the first line is the exact ``experimental likelihood'' and the second line is the SL. Here $\pi(\bm \delta)$ is the joint probability density distribution for the elementary nuisance parameters. In our assumptions  these are independent from each other, hence the prior factorises as $\pi(\bm \delta)=\prod_{i=1}^N \pi_i(\delta_i)$. The SL formalism shown here is relevant for $N\geq P$, which is also the most common case.%
\footnote{If $P<N$, there are more observed quantities than nuisance parameters.
 In such case, using the SL at the level of the event rates, although not formally wrong, is inappropriate.
  Equation~\eqref{eq:SL_master} still applies but the covariance matrix will be singular. In the case of Higgs theoretical uncertainties for example, a more appropriate combination is done at the level of cross-sections and branching ratios, as realised in~\cite{Arbey:2016kqi}. Another example is the one of unbinned likelihoods, for which parametric functions for the signal and background probability densities
are typically used to construct the experimental likelihood. The systematic uncertainties are then on the parameters of the signal and background functions. 
Notice that in such case, the shapes can be directly provided in their analytic form by the experimental collaborations.}
The derivation is shown in Sec.~\ref{se:SL_theory}.

The coefficients $a_I$, $b_I$ and $c_I$, and the $P\times P$ correlation matrix $\bm{\rho}=\rho_{IJ}$ define the SL and are in general functions of the parameters of interest. However, in concrete cases, this dependence will often be negligible. This is in particular the case in particle physics searches for new physics when the expected event number decomposes into signal ($n_s$) plus background ($n_b$) contributions. The parameters of interest that model the new physics enter in $n_s$ while $n_b$ is independent of them.  Whenever the expected signal is small with respect to the background, the dominant uncertainties in searches for new physics are those related to the background.
Neglecting the systematic uncertainties affecting the signal implies in turn that the parameters of the SL are independent of $\bm \alpha$. 
Hence the SL Eq.~\eqref{eq:SL_master} takes the form\,\footnote{We have substituted
$a_I(\alpha) \equiv a_I+n_{s,I}(\alpha),~b_I(\alpha) \equiv b_I$ and $c_I(\alpha)\equiv c_I$.}
\begin{equation}
 L_{\rm{S}}(\bm{\alpha},\bm{\theta})=
\prod_{I=1}^P \mathrm{Pr}\Big( n^{\rm obs}_I \, \Big| \, n_{s,I}(\bm{\alpha})+a_I+ b_I\theta_I+c_I\theta_I^2  \Big) \cdot
\frac{ \mathrm{e}^{ \textstyle-\frac{1}{2}\bm{\theta}^\mathrm{T} \bm{\rho}^{-1} \bm{\theta} }}{\sqrt{(2\pi)^P}},
\label{eq:SL_LHC}
\end{equation}
which is the expression we use in the rest of this paper. 
(Non-negligible signal uncertainties will be commented on in Sec.~\ref{se:application}.)
The expression Eq.~\eqref{eq:SL_LHC} is valid for data with \textit{any statistics of observation}. 
Since the data in particle physics are often observed event counts, $n_{I}^{\rm{obs}}$, they will typically follow Poisson statistics such that
\be
\textrm{Pr}(n^{\rm obs}_{I}|n_{I})\equiv \textrm{Pois}({n}^{\rm{obs}}_{I}|n_{I}) = \dfrac{(n_{I})^{{n}^{\rm{obs}}_{I}} \mathrm{e}^{-n_{I}}}{{n}^{\rm{obs}}_{I}!} \, .
\ee
However, as mentioned, the formalism presented here applies regardless of the dependence on the parameters of interest. For example, the likelihood can very well be multimodal in the parameters of interest; this does not affect the validity of the  approach.

The parameters of the SL ($a_I, b_I, c_I, \rho_{IJ}$) have analytical expressions
as a function of the variance and the skew  of each elementary nuisance parameter (see Section~\ref{se:analytic}). However, often the elementary uncertainties and the event yields are already coded in a Monte Carlo (MC) generator. In this case, an elegant method to obtain the SL parameters  is the following. From the estimators of the event yields $\hat n_I$, one can evaluate the three first moments of the $\hat n_I$ distribution and deduce the parameters of the SL directly from these moments.
What is needed is the mean $m_{1,I}$, the covariance matrix $m_{2,IJ}$ and the diagonal component of the third moment $m_{3,I} \equiv m_{3,III}$. 

Using the definition $ n_I= a_I+b_I \theta_I+c_I \theta_I^2 $, we have the relations
\begin{align} \label{eq:moments1}
  m_{1,I} &= \mathbf{E}[\hat n_I]= a_I+c_I\,,\\
  m_{2,IJ} &= \mathbf{E}[(\hat n_I - \mathbf{E}[\hat n_I])(\hat n_J - \mathbf{E}[\hat n_J]) ]= b_I b_J \rho_{IJ}+2 c_I c_J\rho_{IJ}^2\,,\\
  m_{3,I} &= \mathbf{E}[(\hat n_I - \mathbf{E}[\hat n_I])^3 ]=   6 b_I^2 c_I+8 c_I^3 \, ,
  \label{eq:moments2}
\end{align}
where $\mathbf{E}$ denotes the expectation value.
Inverting these relations, while taking care to pick the relevant solutions to quadratic and cubic equations, gives the parameters of the SL.  We find
\begin{align}
c_I &= \, -\mathrm{sign}(m_{3,I}) \, \sqrt{2 m_{2,II}} \, \cos\!\left(\frac{4\pi}{3}+\frac{1}{3}\arctan\left(\sqrt{8 \frac{m^3_{2,II}}{m^2_{3,I}}-1}\right)  \right)\, , \label{eq:solutions1}
\\
b_I &= \, \sqrt{m_{2,II}-2 c_I^2 }\,,\\
a_I &= \, m_{1,I}- c_I\,,\\
\rho_{IJ} &= \, \frac{1}{4 c_I c_J}\left( \sqrt{(b_I b_J)^2+8 c_I c_J\,m_{2,IJ}}-b_I b_J \right) \, .
\label{eq:solutions2}
\end{align}
These formulae apply if the condition $8 m_{2,II}^3\geq m_{3,I}^2$ is satisfied.   
Near this limit,  the asymmetry becomes large and the approximation inaccurate  
because higher order terms $O(\theta_I^3)$  would need to be included in Eq.~\eqref{eq:SL_master}. 
In practice, however, this requires a high skewness of the nuisance parameters, and the SL framework up to quadratic order is sufficient for most applications.

This method will be used in the examples shown in the rest of the paper. 
 This means that if one is provided with the moments $m_{1}$ and $m_{3}$ for each bin and the covariance matrix $m_{2,IJ}$, the SL parameters are completely defined. Moreover, in the case where the nuisance parameters affect only the background rate Eq.~\eqref{eq:SL_LHC}, this computation has to be realised only once and the resulting likelihood can be used for any kind of signal by appropriate substitution of $n_{s}(\bm{\alpha})$.

A reference code implementing the SL and subsequent test statistics is described in Appendix~\ref{app:reference_code}
and publicly available at \url{https://gitlab.cern.ch/SimplifiedLikelihood/SLtools}.

\section{The simplified likelihood from the central limit theorem}
\label{se:SL_theory}

This section contains the derivation of the SL formula Eq.~\eqref{eq:SL_master}.
The reader interested only in the practical aspects of the SL framework can safely skip it. In Section~\ref{se:skew_CLT} we lay down a  result about the next-to-leading term of the CLT. In Section~\ref{se:analytic} we then demonstrate Eq.~\eqref{eq:SL_master} and give the analytical expressions of the SL parameters as a function of the elementary uncertainties. The precision of the expansion is discussed in Section~\ref{se:precision}. 

\subsection{Asymmetries and CLT at next-to-leading order}
\label{se:skew_CLT}

In the classical proof of the CLT, a Taylor expansion is applied to the characteristic functions of the random variables.
Within this Taylor expansion, usually only the leading term is considered, resulting in an asymptotically Normal behavior for
the sum of the random variables.
In the context of the SL framework, however, the next-to-leading  term in the CLT's large $N$ expansion is also considered. 
 This next-to-leading term encodes skewness, which encodes information about the asymmetry of the distribution. This asymmetry is a relevant feature for the analyses hence it is in principle safer to keep this information. 
Another reason to take the asymmetry  into account is that event yields are defined on $\mathbf{R}^+$, while the normal distribution takes values on 
 $\mathbf{R}$. Thus, keeping only the leading order distribution can lead to negative yields.
Such unphysical results can be interpreted as an indicator that the leading order approximation (namely the normal distribution) is too inaccurate. When taking the next-to-leading term into account, an asymmetric support -- such as $\mathbf{R}^+$ --  becomes possible, such that the issue of negative yields disappears. Concrete examples of this feature will be shown in  Fig.~\ref{fig:approxs}. 
\footnote{It is in principle possible to truncate the Gaussian prior by requiring that the expected background plus signal be positive.  However in the presence of signal uncertainties with truncated Gaussian prior, the posterior can become improper   (see \textit{e.g.}~\cite{Obj_Bayes}). This can be understood as a pathology of such approach. In contrast, the alternative we propose does not require truncation. }

The deformed Gaussian obtained when keeping the skew into account does not seem to have in general an analytical PDF. 
However, by using the large $N$ expansion, we are able to express the CLT at next-to-leading order in a very simple way. 
We realise that a random variable $Z$ with characteristic function
\be
\varphi_Z(t)=\exp\left(-\frac{\sigma^2 t^2}{2}-i \frac{\gamma t^3}{6 \sqrt{N}} +O\left(\frac{t^4}{N}\right)\right) \label{eq:CF_CLT}
 \ee
can, up to higher order terms in the large $N$ expansion, be equivalently be expressed in terms of an exactly Gaussian variable $\theta$ in the form
\be
Z= \theta+\frac{\gamma}{3\sqrt{N}}\theta^2\,,  \quad \textrm{with } \quad \theta\sim{\cal N}(0,\sigma^2)\,. \label{eq:ZskewCLT}
\ee
We will refer to this type of expression as ``normal expansion''. Details about its derivation are given in Appendix~\ref{app:skew}.

Equation~\eqref{eq:ZskewCLT}  readily gives the most basic CLT at next-to-leading order when assuming $Z=N^{-1/2}\sum_{j=1}^N \delta_j$, where the $\delta_j$ are  independent identically distributed centred nuisance parameters  of variance $\sigma^2$ and third moment $\gamma$. The method works similarly with the Lyapunov CLT, \ie\ when the $\delta_j$ are not identical and have different moments $\sigma^2_j$, $\gamma_j$, in which case one has defined $\sigma^2=N^{-1}\sum_{j=1}^N \sigma_j^2$, $\gamma=N^{-1}\sum_{j=1}^N \gamma_j$,

Finally, our approach applies similarly to the multidimensional case where various linear combinations of the $\delta_j$ give rise to various $Z_I$. The $Z_I$ have a covariance matrix $\Sigma_{IJ}$ and a skewness tensor $\gamma_{IJK}={\rm E}[Z_IZ_JZ_K]$. For our purposes, we neglect the non-diagonal elements of $\gamma$, keeping only  the diagonal elements, denoted $\gamma_{III}\equiv \gamma_I$. These diagonal elements encode the leading information about asymmetry, while the non-diagonal ones contain subleading information about asymmetry and correlations. With this approximation, we obtain the multidimensional CLT at next-to-leading order,
\be
Z_I\rightarrow \theta_I+\frac{\gamma_{I}}{3\sqrt{N}}\theta_I^2\,,\,\,N\rightarrow \infty  \quad \textrm{with } \quad \theta_I\sim{\cal N}(0,\Sigma)\,. \label{eq:NLCFT}
\ee
 This  result will be used in the following. Again, for $\gamma_I\rightarrow 0$, one recovers the standard multivariate CLT.

\subsection{Calculation of the simplified likelihood}
\label{se:analytic}

Let us now prove Eq.~\eqref{eq:SL_master}.  The dependence on the parameters of interest $\bm{\alpha}$ is left implicit in this section. We will first perform a step of  propagation of the uncertainties, then a step of combination. This is a generalisation of the approach of \cite{Fichet:2016gvx}. Here we take into account the skew, hence there is no need to use an exponential parameterisation like in \cite{Fichet:2016gvx}.

In this section the elementary nuisance parameters $\delta_i$ are independent, centered, have unit variance, and have skew $\gamma_i$, \ie
\be
{\bf E}[\delta_i]=0\,,\quad {\bf E}[\delta_i^2]=1 \,,\quad {\bf E}[\delta_i^3]=\gamma_i\,.
\ee
It is  convenient to use a vector notation for the set of these elementary nuisance parameters, $(\delta_i)\equiv \boldsymbol{\delta} $.

As a first step, we want to propagate the systematic uncertainties at the level of the event numbers.  For an event number $n$ depending on a quantity $Q$ subject to uncertainty, we have
\be n[Q]\equiv n[Q_0(1+\Delta_Q \delta)]\,.
\ee
The propagation amounts to performing a Taylor expansion with respect to $\Delta_Q $. This expansion should be truncated appropriately  to retain the leading effects of the systematic uncertainties in the likelihood. It was shown in \cite{Fichet:2016gvx} that  the expansion should be truncated above second order.

For multiple sources of uncertainty, we have a vector $\boldsymbol{\delta}$ and the relative uncertainties propagated to $n$ are written as
\be n\equiv n^{0}\left( 1+ \Delta_{1}^T\cdot \, \boldsymbol{\delta}+\boldsymbol{\delta}^{\rm T}\cdot\Delta_{2} \cdot \boldsymbol{\delta} +O\left(\frac{n^{(3)}}{n^{0}}\Delta_Q^3\right)
\right)\,
\label{eq:propmult}
\ee
with
\be
\Delta_1=\frac{1}{n^{0}}\left(\frac{\partial n }{\partial \delta_{1}}\Delta_{Q,1},\ldots,
\frac{ \partial n }{\partial \delta_{p}}\Delta_{Q,p} \right)_{\boldsymbol{\delta}=0}^{\rm T} \,, \label{eq:stdDelta1} \quad \Delta_2=\frac{1}{2n^{0}}\left(\frac{\partial^2 n}{\partial \delta_{i}\partial \delta_{j}}\Delta_{Q,i}\Delta_{Q,j}  \right)_{\boldsymbol{\delta}=0} 
\ee
and the $n^{(3)}$ denoting schematically the third derivatives of $n$.

The second step is to combine the elementary nuisance parameters. We introduce combined nuisance parameters $\theta_I$ which  are chosen to be centred and with unit variance without loss of generality, and whose correlation matrix  is denoted $\rho_{IJ}$,\ie
\be
 {\bf E}[\theta_I]=0\,, \quad{\bf E}[\theta^2_I]=1\,,\quad{\bf E}[\theta_I \theta_J]=\rho_{IJ}\,.
\ee
Moreover we define the expected event number in terms of the combined nuisance parameters as
\be
n_I=n^{0}_{I}(1+\Delta_{1,I}\cdot\boldsymbol{\delta}+\boldsymbol{\delta}\cdot\Delta_{2,I}\cdot\boldsymbol{\delta}) \equiv
a_I+b_{I}\theta_I+ c_I\theta_I^2 \label{eq:comb_def}\,.
\ee
The $a_I, b_I, c_I$ parameters together with  the correlation matrix $\rho_{IJ}$ fully describe the combined effect of the elementary uncertainties.
To determine them we shall identify the three first moments on each side of Eq.~\eqref{eq:comb_def}. We obtain
 \be
a_{I}=n^{0}_{I}\left(
1+\tr \Delta_{2,I}-\frac{1}{6}\sum_{i=1}^N \gamma_i(\Delta_{1,I,i})^3+O(\Delta^4) \right) \label{eq:n_comb}
\,,
\ee
\be
b_I=a_I\left(\Delta_{1,I}^{\rm T}.\Delta_{1,I}  +2\sum_{i=1}^N \gamma_i\Delta_{1,I,i}\Delta_{2,I,i}+O(\Delta^4)\right)^{1/2}\,,
\ee
\be
\rho_{IJ}= \frac{a_I a_J}{b_I b_J}\left(\Delta_{1,I}^{\rm T}.\Delta_{1,J}  +\sum_{i=1}^N \gamma_i(\Delta_{1,I,i}\Delta_{2,J,i}+\Delta_{1,J,i}\Delta_{2,I,i})\right)+O(\Delta^4)
\,,
\ee
\be
c_I=\frac{a_I}{6}\sum_{i=1}^N \gamma_i(\Delta_{1,i})^3+O(\Delta^4) \label{eq:gam_comb}\,,
\ee
where the $O(\Delta^4)$ denotes higher order terms like $\tr(\Delta_{2,I}^{\rm T}\cdot\Delta_{2,I})$, $(\tr\Delta_{2,I})^2$, $\Delta_{1,I}^{\rm T}\cdot\Delta_{1,I}\tr\Delta_{2,I}$ which are neglected.
When $\gamma_i\rightarrow 0$  one recovers the expressions obtained in Ref.~\cite{Fichet:2016gvx}.\footnote{For simplicity we show here the expressions assuming $c_I\ll b_I$, as it is sufficient in the scope of the proof. For sizeable $c_I$, one should instead use the exact solutions of the system, Eqs.~\eqref{eq:solutions1}--\eqref{eq:solutions2}.}

Importantly, the $\Delta_2$ term contributes at leading order only in the mean value $a_I$ and always gives 
subleading contributions to higher moments. Hence, for considerations on higher moments, which define the  
shape of the combined distribution, we can safely take the approximation
\be
n_{I}\approx n^{0}_{I}\left(1+ \Delta_{1,I}\cdot \boldsymbol{\delta}\right)\, \label{eq:n_approx}
\ee
from Eq.~\eqref{eq:comb_def}.
We now make  the key observation that this quantity is the sum of a large number of independent random variables. These are exactly the conditions for a central limit theorem to apply.
As all the elementary uncertainties have in principle different shape and magnitudes we apply Lyapunov's CLT \cite{Billingsley}. We can for instance use Lyapunov's condition on the third moment, and the theorem reads as follows. If 
\be
\quad\frac{{\bf E}[(n_I-{\bf E}[n_I])^3]}{{\bf E}[(n_I-{\bf E}[n_I])^2]^{3/2}}\sim\frac{6 c_I}{b_I}\rightarrow 0 \quad \,\textrm{for} \quad N\rightarrow \infty
\ee
then
\be
\theta_I\sim {\cal N }(0,\rho) \quad \,\textrm{for} \quad N\rightarrow \infty \,.
\ee

Furthermore we can see that the expression of $n_I$ in terms of the combined nuisance parameters, $n_I=a_I+b_{I}\theta_I+ c_I\theta_I^2$ (first defined in Eq.~\eqref{eq:comb_def}), takes the form of a normal expansion, see Eq.~\eqref{eq:NLCFT}. 
This means that the $c_I \theta_I^2$ term  corresponds  precisely to the leading deformation  described by the next-to-leading term of the CLT.  This deformation encodes the skewness induced by the asymmetric elementary uncertainties. We have therefore obtained  a  description of the main collective effects of asymmetric elementary uncertainties, which is dictated by  the CLT. The resulting  simplified likelihood is given in Eq.~\eqref{eq:SL_master}.

\subsection{Precision of the normal expansion}
\label{se:precision}

The accuracy of the normal expansion $n =  a+ b \theta+ c \theta^2$ with $\theta\sim {\cal N}(0,1)$ --- and thus of the  simplified likelihood --- is expected to drop when only a few elementary  uncertainties are present and these depart substantially from the Gaussian shape. 
This is the situation in which the next-to-leading CLT, Eq.~\eqref{eq:NLCFT}, tends to fail.  It is instructive to check  on a simple case how the normal expansion approximates the true distribution, and in which way discrepancies tend to appear.

We consider the realistic  case of a log-normal distribution with parameters $\mu, \sigma$. We fix $\mu=0$ without loss of generality. The three first centered moments are
\be
m_1=e^{\frac{\sigma^2}{2}}\,,\quad m_2=e^{2\sigma^2}-e^{\sigma^2}\,,\quad
m_3=e^{\frac{9\sigma^2}{2}}-3 e^{\frac{5\sigma^2}{2}}+2 e^{\frac{3\sigma^2}{2}}
\ee
and  $a,b,c$ are obtained  using  Eqs.~\eqref{eq:solutions1}--\eqref{eq:solutions2}.

For $\sigma\sim 0.69$, the bound $8m_2^3\approx m_3^2$  is reached (see Section~\ref{se:EL_SL}). This is the limit where
the distribution is so asymmetric that the variance comes entirely from the $\theta^2$ term. Beyond this bound the  normal expansion cannot be used at all as Eqs.~\eqref{eq:solutions1}--\eqref{eq:solutions2} have no solutions.
The distribution has $c>0$ thus $n$ has a lower bound given by $n>a-b^2/4c$. 
Below this limit on $\sigma$, the lower bound on $n$ is roughly $n\gtrsim 0.5$, therefore the approximation can never produce a negative event yield.

\begin{figure}[t]
\begin{center}
\includegraphics[width=0.7\textwidth]{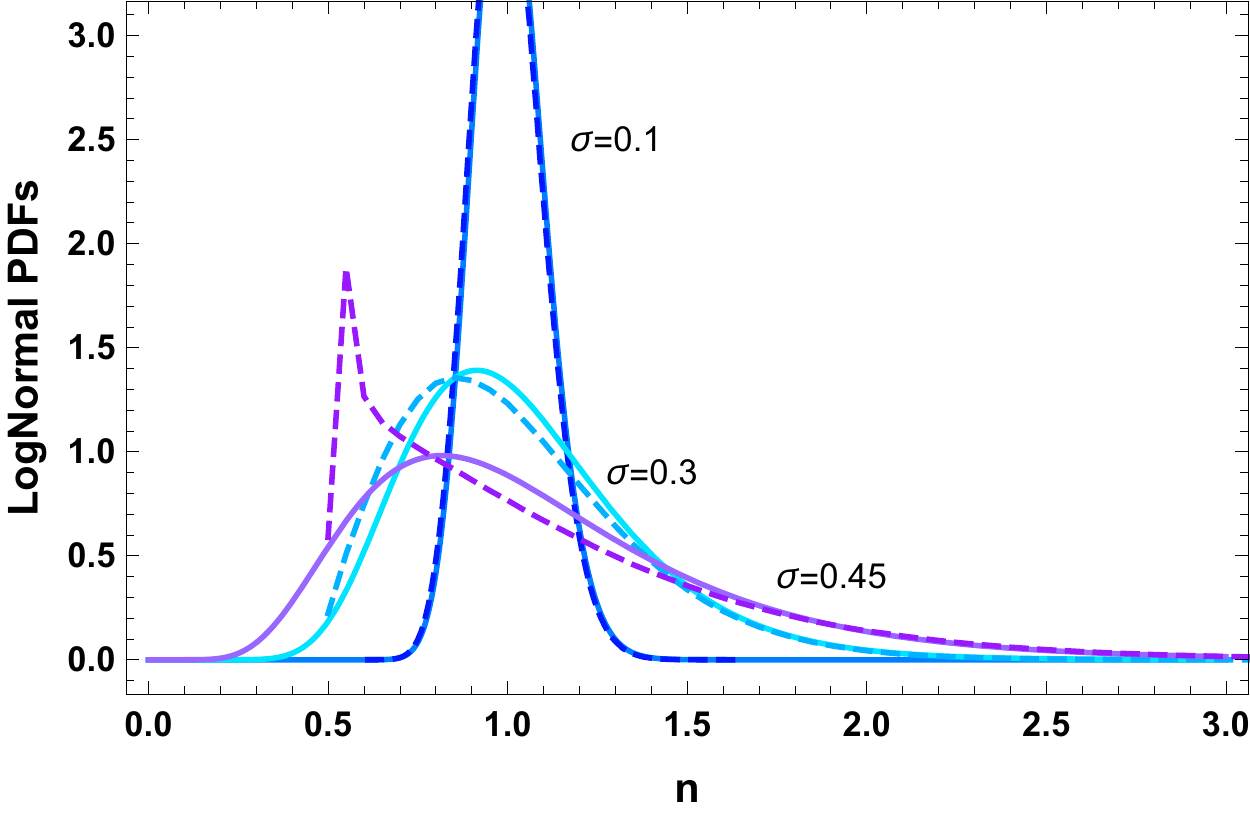}
\end{center}
\caption{\label{fig:approxs} 
The log normal PDFs and corresponding normal approximations for $\sigma = 0.1$, 0.3 and 0.45 are shown in blue, cyan and purple respectively. Solid curves show the true distributions, dashed curves show the approximate distributions. 
}
\end{figure}

To check numerically how well the approximation performs, the true and approximate PDFs are compared in Figure~\ref{fig:approxs} for various values of $\sigma$.  
Since the approximate PDF never gives $n<0.5$, it can only be a good approximation if the true PDF is vanishing in this region. This is the case for
asymmetries $\sigma \lesssim 0.3$, and as can be seen in the figure the normal approximation indeed works very well.
For larger asymmetries, $\sigma = 0.45$ in our example, the true PDF becomes sizeable in the region $n<0.5$. 
The approximation still performs reasonably well for larger $n$, however, near $n\sim 0.5$, 
the approximate PDF tends to increase and become peaked to account for the area at $n<0.5$ that it cannot reproduce. 
This behaviour will also be observed for certain bins in the LHC-like analysis implemented in Sec.~\ref{se:SL_LHC}.

Overall, through this example, we can see that the normal approximation tends to become inaccurate for a skewness of $\sim 100$--$150\%$. 
This is a moderate value, however one should keep in mind that these considerations apply to the combined uncertainties, for which small skewness is typical. The accuracy of the SL framework will be tested in a realistic setup in Sec.~\ref{se:SL_LHC}.

\section{Practical aspects of the simplified likelihood framework}
\label{se:practice}

\subsection{Range of application}
\label{se:application}
An important feature of the SL is that it is flexible in the sense that the combination of the systematic uncertainties does not have to be applied to the whole set.  The only requirement to combine a subset of the  uncertainties is that it should have a convergent enough  CLT behaviour in order  for the SL to be accurate. There is thus a freedom in partitioning the set of systematic uncertainties, giving rise to variants of the SL that can be either equivalent or slightly different upon  marginalising.

For instance, if a single systematic uncertainty $\delta$ is left apart from the combination, the SL takes the form
\be
 L_{\rm{S}}(\bm{\alpha},\bm{\theta}) = \prod_{I=1}^P \mathrm{Pr}\Big( \hat{n}_I\,\Big|\,a_{I}(\bm{\alpha})+b_I\theta_I+c_I\theta_I^2 + \Delta_I  \delta   \Big) \cdot
    \frac{ \mathrm{e}^{{\textstyle-\frac{1}{2}\bm \theta^{\rm T} \bm{\rho}^{-1} \bm \theta}}}{\sqrt{(2\pi)^P }}
    \cdot \pi(\delta) \,. \label{eq:SL_variants1}
\ee
Similarly, if two subsets of systematic uncertainties $\bm{\theta}$ and $\bm{\tilde \theta}$ tend to  separately satisfy the CLT condition,  they can be separately combined, giving  
\be
 L_{\rm{S}}(\bm{\alpha},\bm{\theta},\bm{\tilde \theta}) = \prod_{I=1}^P \mathrm{Pr}\Big( \hat{n}_I\,\Big|\,a_{I}(\bm{\alpha})+b_I\theta_I+c_I\theta_I^2 + \tilde b_I \tilde \theta_I+ \tilde c_I \tilde \theta_I^2  \Big) \cdot
    \frac{ \mathrm{e}^{{\textstyle-\frac{1}{2}\bm \theta^{\rm T} \bm{\rho}^{-1} \bm \theta}}}{\sqrt{(2\pi)^P }}
\cdot \frac{ \mathrm{e}^{{\textstyle-\frac{1}{2}\bm{ \tilde\theta^{\rm T}} \bm{\tilde \rho}^{-1} \bm{ \tilde\theta}}}}{\sqrt{(2\pi)^P }}
      \,. \label{eq:SL_variants2}
\ee

The SL naturally accommodates any such partitions. It is actually  commonplace in LHC analyses   to present systematic uncertainties combined in subsets, for example ``theoretical'', ``experimental'', ``luminosity'', ``MC'' uncertainties. 
This is useful not only for informative purpose but also for further interpretations. 
For example the theoretical uncertainties may be improved later on 
and it is clearly of advantage if their effect can be re-evaluated without having to re-analyse the whole data (which could only be done by collaboration insiders).\footnote{Such combination of theoretical uncertainties  has been done in \cite{Arbey:2016kqi}  for the Higgs production and decay rates and can be implemented in a Higgs SL.}
Another reason to single out a nuisance parameter  from the combination (as shown in Eq.~\eqref{eq:SL_variants1})
is if it has a large non-Gaussian PDF that one prefers to take into account exactly.
In order to profit from the  versatility of the SL, an equally  versatile format is needed to release the SL data. This will be the topic of next subsection.

Finally, some considerations are in order regarding  \textit{signal uncertainties}. 
The expected rate $n$ given in Eq.~\eqref{eq:propmult} splits as $n=s+b$ where $s$ is the signal and $b$ the background. 
Each elementary nuisance parameter $\delta_i$ can in principle affect both $s$ and $b$. 
The most general form taken by the expected rate is then
\begin{align} n=s+b & \equiv s^{0} \left( 1+ \Delta_{1,s}^T\cdot \, \boldsymbol{\delta}+\boldsymbol{\delta}^{\rm T}\cdot\Delta_{2,s} \cdot \boldsymbol{\delta} \right) \nonumber
+ b^{0} \left( 1+ \Delta_{1,b}^T\cdot \, \boldsymbol{\delta}+\boldsymbol{\delta}^{\rm T}\cdot\Delta_{2,b} \cdot \boldsymbol{\delta} \right) \\
 & = (s^{0}+b^0) \left( 1+ \Delta_{1}^T\cdot \, \boldsymbol{\delta}+\boldsymbol{\delta}^{\rm T}\cdot\Delta_{2} \cdot \boldsymbol{\delta} \right)\,
\label{eq:propmultSB}
\end{align}
with
\be
\Delta_{1}= \frac{s^{0}\Delta_{1}^s+b^0 \Delta_{1}^b}{s^{0}+b^0}\,,\quad
\Delta_{2}= \frac{s^{0}\Delta_{2}^s+b^0 \Delta_{2}^b}{s^{0}+b^0}\,.
\label{eq:Deltas}
\ee
The $\Delta^s_{1}$, $\Delta^s_{2}$ vectors encode the contributions from the signal, while the $\Delta^b_{1}$, $\Delta^b_{2}$ vectors encode the contributions to the background. 
The signal $s^{0}$ and possibly $\Delta^s_{1}$, $\Delta^s_{2}$ depend on the parameters of interest ${\bm \alpha}$. 
For discovery or limit-setting, the uncertainties on the signal are a subleading effect. In this paper, as said in Sec.~\ref{se:EL_SL}, 
we have neglected signal uncertainties ($\Delta_{1}^s=\Delta_{2}^s=0$).
In this approximation, only the background uncertainties remain in Eq.~\eqref{eq:propmultSB}, 
and thus the SL does not depend on ${\bm \alpha}$. 
A similar discussion of the signal$+$background case and a toy-model testing the SL in this case has been done in \cite{Fichet:2016gvx}. 

The inclusion of pure signal uncertainties is fairly straightforward because their contribution factors out from the background ones. 
In \eqref{eq:Deltas}, this means that the vectors $\Delta_{1,2}$ can be simply organised as the union of the subvectors $\Delta_{1,2}=(\Delta^b_{1,2},\Delta^s_{1,2})$. 
This implies that the pure signal uncertainties do not affect the SL parameters $a$, $b$, $c$, $\rho$, and can thus be rigorously included directly within the existing SL (for $\Delta_{1}^s=\Delta_{2}^s=0$).

In contrast, for correlated systematic uncertainties affecting both signal and background --- for instance in case of measurements as opposed to limit setting, or when interference effects between signal and background are important --- 
the $b,c,\rho$ parameters  become dependent on the parameters of interest ${\bm \alpha}$.  
This requires to (re-)derive the SL taking into account all the elementary nuisance parameters at once, which is a much heavier task. 

Altogether, while there is no conceptual difference regarding the SL formalism with or without signal uncertainties, there are important practical implications. Numerical evaluations become much heavier when the parameters of the SL---especially $\rho_{IJ}(\alpha)$ which requires a matrix inversion---have to be evaluated for each value of ${\bm \alpha}$.  The presentation of the SL data, iscussed in the next subsection, may also become more evolved.  
Furthermore, and perhaps most importantly, the SL is then valid only for the particular signal assumption it has been derived for. 


\subsection{Construction and  presentation}

There are in principle two ways of releasing the data needed to build the simplified likelihood. One way 
is to release the whole set of elementary systematic uncertainties, the other to release the three first moments of the PDF of the combined systematic uncertainties. While the former is in principle doable, we will focus only on the latter. Indeed, the elementary uncertainties are usually already  coded by the experimentalists in MC generators, hence it is straightforward to evaluate these moments.
\footnote{Using the elementary uncertainties maybe more convenient when one wishes to include the systematic uncertainties on the signal, \ie\  $\alpha$-dependent $b,c,\rho$. Since these systematics are not crucial for new physics searches we do not take them into account here.}

We thus focus on the release of the  SL data via the  $m_{1,I}$, $m_{2,IJ}$, $m_{3,I}$
moments   of the PDF of the combined systematic uncertainties,  already defined in Eqs.~\eqref{eq:moments1}--\eqref{eq:moments2}, where $m_{3,I}$ is the diagonal part of the third-rank tensor $m_{3,IJK}$.  Evaluating these moments via MC toys is straightforward for the experimental analysis. However, their way of presentation needs to be  
considered in detail, taking into account  the available tools and the current practices. This is the purpose of this subsection.

Key to the usefulness of any likelihood data for analysis reinterpretation is
the availability of that data in a standard format. For global fits, where tens
or hundreds of analyses may be used simultaneously, it is crucial that this
format be unambiguously parseable by algorithms without human assistance. A
standard location is also necessary, for which the obvious choice is the longstanding
HEP data repository, HepData~\cite{hepdata}.

It is convenient to refer to the data in terms of
the order of the moment from which they originate. We will use the term ``$n$-th
order data'' to refer to information coming from a moment of order $n$; here, $n$
will go only up to $3$. Second-order data includes the covariance matrix,
correlation matrix, and/or diagonal uncertainties: these can be given either in a
relative or absolute parametrisation. There is the same kind of freedom for
third-order data but this does not need to be discussed here. In addition to the
moments of the combined systematic uncertainties, this terminology will also
apply to the observed central values and statistical uncertainties usually
presented by the experiments.

Let us review the current formats of presentation of likelihood data. The presentation
of first-order data is standardised while currently no third-order data are usually given.
Regarding second-order data there is unfortunately no standard representation
currently established.  A review of the second-order data in HepData 
and on the experiments' analysis websites reveals a
mixture of presentation styles:
\begin{itemize}
\item \textit{Table format:} 2D histograms of either covariance or correlation
  matrices. This has the difficulty that the convention used is not made clear
  (other than by inspection of the matrix diagonal), and without a structural
  association with a first order dataset it is impossible for computer codes to
  unambiguously construct the relevant likelihood. In the case of the
  presentation of a correlation (as opposed to covariance) matrix, the diagonal
  variances must be provided with the first-order dataset.

\item \textit{Error source format:} A vector of labeled $\pm$ terms associated
  to each element of the first-order dataset. The correlations between the error
  sources is indicated via the labels, (\eg, a ''\texttt{stat}'' label to be a
  purely diagonal contribution, a ``\texttt{lumi}'' label to be 100\% correlated
  across all bins, and all other labeled uncertainties treated as orthogonal).
  The correlation or covariance matrices can be constructed using Eq.~\eqref{eq:sum_errsource}.
  This format presents the second-order data in the form of ``effective''
  elementary uncertainties.

\item Auxiliary files in arbitrary format: the \emph{ad hoc} nature of these
  makes them impossible to be handled by unsupervised algorithms. This includes
  2D histograms in ROOT data files, since variations in path structure and the
  ambiguity between covariance or correlation matrices are an impediment to
  automated use. This presentation style will be disregarded below.
\end{itemize}
The table and error source formats  may be readily extended for automated data
handling and are thus appropriate to release SL data.

In the case of the table format, in addition to the observed central values and
statistical uncertainties usually released, extra HepData tables can encode the
$m_{1,I}$, $m_{2,IJ}$, $m_{3,I}$ moments describing the combined nuisance
parameters.  However the HepData table headers will have to be augmented in a
standardised fashion to express the relationships between tables,
\ie\ unambiguously identifying the moment data tables associated with a
first-order dataset. While the format is conceptually straightforward,
introducing the semantic description of the tables is at present highly
impractical. We hence recommend the error source format for which identifying
the associations between datasets is trivial.

In the error source format, the $m_{1,I}$, $m_{2,IJ}$, $m_{3,I}$ moments are
\textit{all} encoded in the form of labeled vectors. The $m_{2,IJ}$ matrix is
reconstructed via a sum of the form \be m_{2,IJ}= \sum {a_{I,i}a_{J,i}} \, 
\label{eq:sum_errsource}
\ee
where the $a_{I,i}$ are the released error sources.
The vector of third order data can be  indicated via a special label.
There is not limit in the number of labels associated to an element hence this format is very flexible.
For instance the $a_{I,i}$ error sources corresponding to the decomposed covariance  can just get bland names such as ``{\tt sys,NP1}'', but  can also be extended with, \eg, a ``{\tt th}'' prefix to allow separation of experimental and theory systematics (since the theory can in principle be improved on in future reinterpretations).

This format requires some keyword standardisation.  The final scheme should
be capable of equally applying to any kind of experimental data and systematic uncertainties. In particular it should be valid for event counts,  differential cross-sections with bins correlated by the systematic uncertainties, correlations between  the bins of different distributions/datasets, and so on.

Summarising, our recommendation is to release the moments of the combined uncertainty distributions via the HepData error source format, which has built-in semantics of arbitrary complexity and can thus make the most of the SL framework. As a showcase example, we provide the pseudo-data used in the next section as a sandbox HepData record at \url{https://www.hepdata.net/record/sandbox/1535641814}.

\section{Simplified likelihood in a  realistic LHC-like analysis }
\label{se:SL_LHC}

In this section we introduce a realistic pseudo-analysis that is representative of a search for new physics at the LHC. 
This analysis will be used to validate the SL method and to test its accuracy in realistic conditions. 
It is also used to validate the SL reference code presented in Appendix~\ref{app:reference_code}. 
Finally, this pseudo-analysis provides a concrete example of SL data release via the HepData table format (see above). 
The SL and subsequent results of the pseudo-search can be reproduced using these data. 

As already mentioned in Section~\ref{se:EL_SL}, the dominant systematic uncertainties relevant in searches for new physics are those related to the background processes. 
Imperfect knowledge of detector effects or approximations used in the underlying theoretical models will lead to uncertainties in the predictions of these processes.
Any mis-estimation of the background could result in an erroneous conclusion regarding the presence (or absence) of a signal.
There are a number of different ways in which an experimentalist may assess the effect of a given systematic uncertainty, but generally, these effects are parameterised using knowledge of how the estimation of a given process which change under variations of some underlying parameter of the simulation model, theory, detector resolution, etc. Estimates of the contribution from background processes are obtained either from simulation or through  data-driven methods.
In the following section, we describe a pseudo-search for new physics, inspired by those performed at the LHC, in which systematic uncertainties are included, and derive the SL parameters for it.

\subsection{A LHC-like pseudo-search for new physics}
\label{se:toy_search}

In order to illustrate the construction of the SL, a model has been constructed which is representative of a search for new physics at the LHC. Typically, in these searches the observed events are binned into histograms in which the ratio of signal to background contribution varies with the bin number. A search performed in this way is typically referred to as a ``shape'' analysis as the difference in the distribution (or shape) of the signal events, compared to that of the background, provides crucial information to identify a potential signal.

\begin{sidewaysfigure}[h!]
\begin{center}
\includegraphics[width=\textwidth]{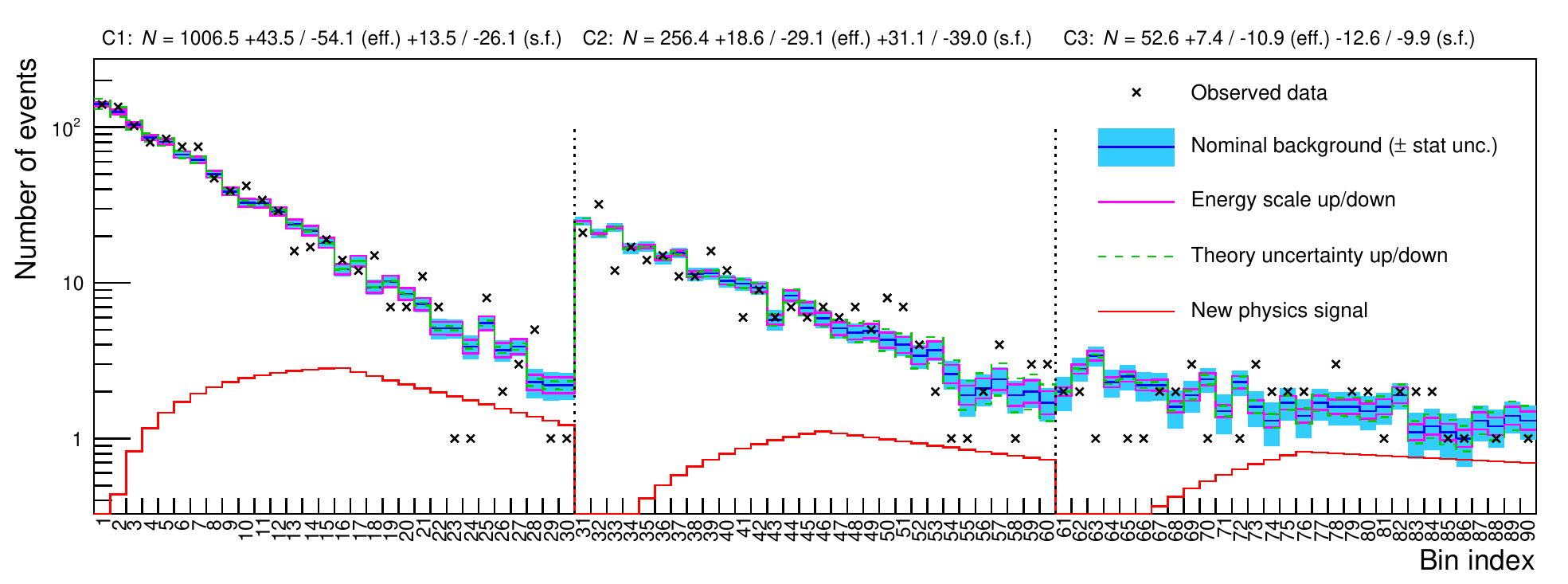}
\end{center}
\caption{LHC-like search for new physics (mockup). The search is performed across three event categories, each divided into 30 bins to make a total of 90 search regions. The nominal expected contribution in each bin from the
background and from the new physics signal is shown by the blue and red lines, respectively. The solid and dashed lines show the $\pm1\sigma$ correlated variation in each bin expected due to an experimental and theoretical
uncertainty while the blue shaded band shows the uncorrelated uncertainty in each bin due to limited MC simulation. The ``observed'' number of events in data in each bin is indicated by the black points.}
\label{fig:toy}
\end{sidewaysfigure}

Our pseudo-search requires to make  assumptions for an ``observed'' dataset, for the corresponding background, and for the new physics signal.  These ingredients are summarised in Figure~\ref{fig:toy},   
which shows the distribution of events, in each of three categories along with the expected contribution from the background 
and the uncertainties thereon, 
and from some new physics signal.
The ``nominal'' background follows a typical exponential distribution where fluctuations are present, representing a scenario in which limited MC  simulation (or limited data in some control sample) was used
to derive the expected background contribution. The uncertainties due to this, indicated by the blue band, are uncorrelated between the different bins. Additionally, there are two uncertainties which modify the ``shape'' of
backgrounds, in a correlated way. The effects of these uncertainties are indicated by alternate distributions representing ``up'' and ``down'' variations of the systematic uncertainty. Finally, there are two uncertainties
which effect only the overall expected rate of the backgrounds. These are indicated in each category as uncertainties on the normalisation $N$ of the background. These uncertainties are correlated between the three categories
and represent two typical experimental uncertainties; a veto efficiency uncertainty (eff.) and the uncertainty from some data-simulation scale-factor (s.f.) which has been applied to the simulation.

\subsection{Parameterisation of backgrounds}

It is typical in experimental searches of this type to classify systematic uncertainties into three broad categories, namely; those which affect only the normalisation of a given process, those which effect both the
``shape'' or ``distribution'' of events of that process in addition to its normalisation, and those which affect only a small number of bins or single bin in the distribution and are largely uncorrelated with the other
bins (eg uncertainties due to limited MC simulation).

The expected (or nominal)\footnote{It should be noted that the expectation value for $n_{b,I}$ is \emph{not} necessarily the same as the mean value. For this reason, we typically refer
to this as the `nominal' value since it is the value attained when the elementary nuisance parameters are equal to their expectation values $\mathbf{\delta}=0$.} number of background events, due to a particular process, in a given bin ($I$) in  
Eq.~\eqref{eq:SL_master} is denoted by
\begin{equation}
  n_{b,I}(\bm{\delta}) \equiv 
  f_{I}(\bm{\delta}) N(\bm{\delta}),
\end{equation}
where the process index ($k$) is suppressed here as we only have a single background process. The functions $N(\bm{\delta})$ and  $f_{I}(\bm{\delta})$ are the total number of expected events for that process in a particular
category and the fraction of those events expected in bin $I$, respectively, for a specified value of $\bm{\delta}$. Often, these functions are not known exactly and some interpolation is performed between known
values of $n_{I}$ at certain values of $\bm{\delta}$. For each uncertainty, $j$, which affect the fractions, $f_{I}$, a number of different interpolation schemes exist. One common method, however, is to interpolate between
three distribution templates representing three values of $\delta_{j}$. Typically, these are for $\delta_{j}=0$, the nominal value, and $\delta_{j}=\pm1$ representing the plus and minus $1\sigma$ variations due to that uncertainty.

The interpolation is given by
\begin{equation}
 f_{I}(\bm{\delta}) = f_{I}^{0}\cdot\frac{1}{F(\bm{\delta})} \prod_{j} p_{Ij}(\delta_{j}),
 \label{eqn:frac_function}
\end{equation}
where $f_{I}^{0}=f_{I}(\bm{\delta}=0)$ and $F(\bm{\delta})=\sum_{I}f_{I}(\bm{\delta})$ ensures that the fractions sum to 1. In our pseudo-search, as there are three event categories,
there are three of these summations, each of which runs over the 30 bins of that category. The polynomial $p_{Ij}(\delta_{j})$ is chosen to be quadratic between values of $-1 \leq \delta_{j} \leq 1$
and linear outside that range such that,
\begin{equation}
 p_{Ij}(\delta_{j}) = \begin{dcases*}
 		\frac{1}{2} \delta_{j}(\delta_{j}-1) \kappa_{Ij}^{-}  -(\delta_{j}-1)(\delta_{j}+1) + \frac{1}{2}\delta_{j}(\delta_{j}+1)\kappa_{Ij}^{+} & for $|\delta_{j}|<1$ \\
        \left[ \frac{1}{2}(3\kappa_{Ij}^{+} + \kappa_{Ij}^{-})-2\right]\delta_{j} - \frac{1}{2}(\kappa_{Ij}^{+}+\kappa_{Ij}^{-})+2 & for $\delta_{j}>1$ \\
         \left[2-\frac{1}{2}(3\kappa_{Ij}^{-} + \kappa_{Ij}^{+})\right]\delta_{j} - \frac{1}{2}(\kappa_{Ij}^{+}+\kappa_{Ij}^{-})+2 & for $\delta_{j}<-1$ \\
    \end{dcases*}
\end{equation}

\clearpage

The values of $\kappa_{Ij}^{-}$ and $\kappa_{Ij}^{+}$ are understood to be determined using the ratios of the template for a $-1\sigma$ variation to the nominal one and the $+1\sigma$
variation to the nominal one, respectively\footnote{The accuracy of this interpolation scheme can be (and frequently is) tested by comparing the interpolation to templates for additional, known values of $f_{I}$ for $\delta_{j}$ values other than $0,-1$ and $1$.}.

For uncertainties which directly modify the expected number of events $n_{i}$ of the distributions, an exponent interpolation is used as the parameterisation.
This is advantageous since the number of events for this process in any given bin is always greater than 0 for any value of $\delta_{j}$. For a relative uncertainty $\epsilon_{Ij}$, the fraction varies as
\begin{equation}
 \frac{n_{b,I}(\bm{\delta})}{n_{b,I}^{0}}  =  \prod_{j} (1+\epsilon_{Ij})^{\delta_{j}}.
  \label{eqn:bin_function}
\end{equation}
This is most common in the scenario where a limited number of MC simulation events are used to determine the value of $n_{b,I}^{0}$
and hence there is an associated uncertainty. As these uncertainties will be uncorrelated between bins of the distributions, most of the terms $\epsilon_{Ij}$ will be 0.

Systematic uncertainties that affect only the overall normalisation are also interpolated using exponent functions,
\begin{equation}
 N(\bm{\delta})  =   N^{0} \cdot \prod_{j} (1+K_{j})^{\delta_{j}},
 \label{eqn:norm_function}
\end{equation}
where $N^{0} = N(\bm{\delta}=0)$ and $j$ runs over the elementary nuisance parameters.  A simple extension to this arises if the uncertainty is ``asymmetric'', as in our pseudo-search;
the value of $K_{j}$ is set to $K^{+}_{j}$ for $\delta_{j}\geq0$ and to $K^{-}_{j}$ for $\delta_{j} < 0$. Furthermore, any uncertainty which affects both the
shape and the normalisation can be incorporated by including terms such as those in Eq.~\eqref{eqn:frac_function} in addition to one of these normalisation terms.
In our pseudo-search, there will be a separate $N(\bm{\delta})$ term for each category which provides the total expected background rate summing over the 30 bins of that category.

Combining Eqs.~\eqref{eqn:frac_function},~\eqref{eqn:bin_function} and~\eqref{eqn:norm_function} yields the full parameterisation,
\begin{equation}
 n_{b,I}(\bm{\delta}) = N^{0}\cdot \prod_{j}(1+K_{j})^{\delta_{j}} \cdot f^{0}_{I} \cdot\frac{1}{F(\bm{\delta})} \prod_{j} p_{Ij}(\delta_{j}) \cdot \prod_{j} (1+\epsilon_{Ij}\delta_{j}).
\label{eqn:expt_param}
\end{equation}

As already mentioned, a typical search for new physics will have contributions from multiple background processes, each with their own associated systematic uncertainties.
Only by summing over all of these backgrounds (\ie\ $n_{b,I}=\sum_{p}n_{b,p,I}$ for different background processes $p$) is the likelihood fully specified.

\subsection{Validation of the simplified likelihood}

Here we compare the true and simplified likelihoods arising from the pseudo-search. It is also instructive to consider the simplified likelihood obtained when neglecting the third moments, \ie\ when setting the coefficients of the quadratic terms $c_I$ to zero in Eq.~\eqref{eq:SL_master}. This less accurate version of the SL will be  referred to as ``symmetric SL'', as opposed to the more precise ``asymmetric SL'' developed in this work.

\begin{figure}[ht]
  \centering
  \includegraphics[width=\textwidth]{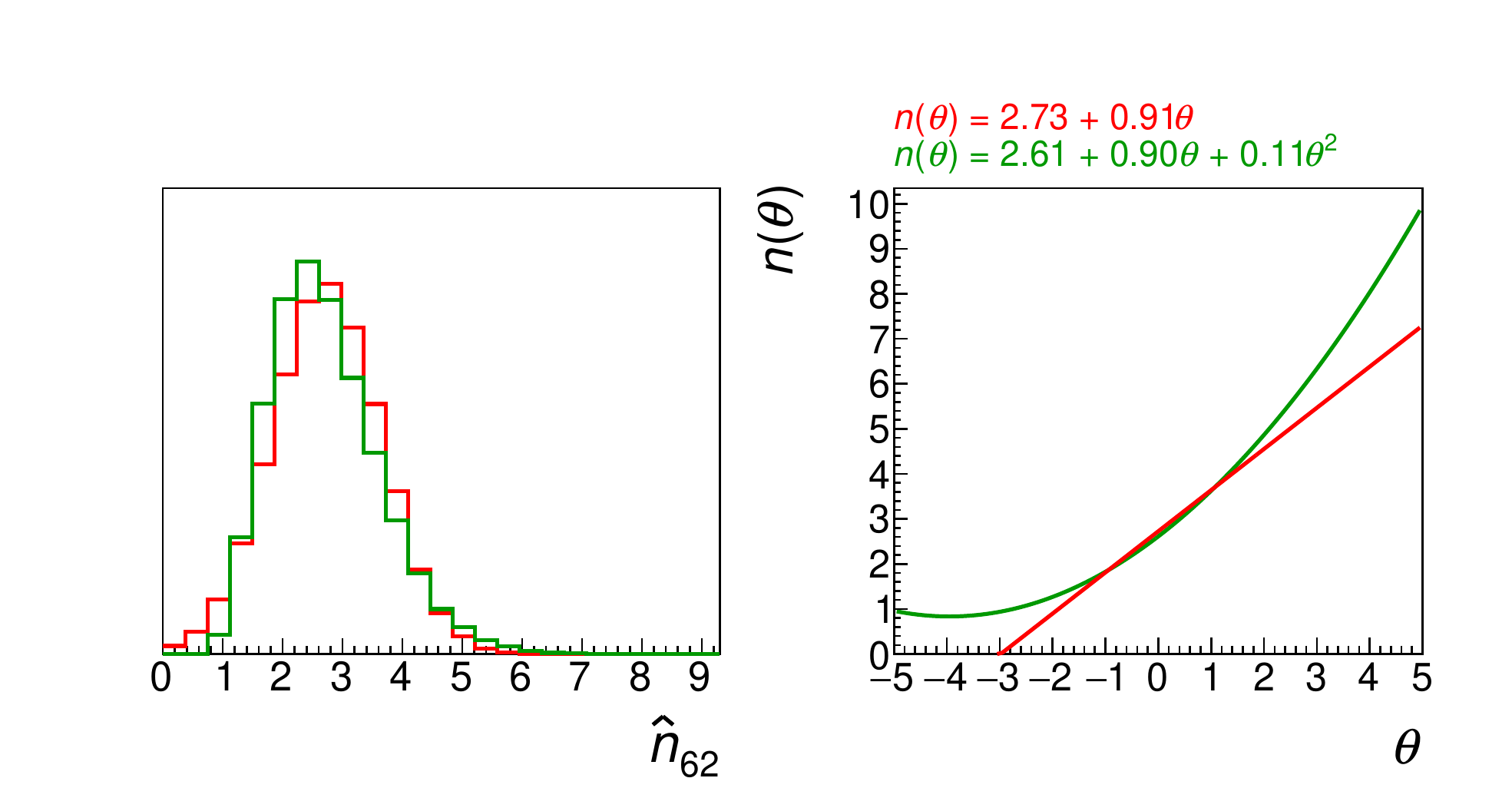}
  \caption{Distributions of $\hat{n}_{I}$ for $I=62$ for the SL.
    The functions $n_{I}(\theta_{I})$ assuming the SL form (green line),
    and when neglecting the third moment (red line), are shown in the right panel while the distributions
    of $\hat{n}_{I}$ obtained for these two cases letting $\hat{\theta}_{I}\sim\mathcal{N}(0,1)$ are shown in the
    left panel.  }
  \label{fig:distributions}
\end{figure}

We constructed 100,000 pseudo-datasets by taking random values ${\hat{\bm{\delta}}}$, generated according to $\pi(\bm{\delta})$, and evaluating
$n_{b,I}(\hat{\bm{\delta}})$ for each dataset according to the Eq.~\eqref{eqn:expt_param}. Figure~\ref{fig:distributions} shows the distribution of $\hat{n}_{i}$, for an example bin, $i=62$,
from the SL. The values of $m_{1},~m_{2}$ and $m_{3}$ are calculated using the pseudo-datasets and subsequently used to calculate the coefficients for the SL.

\begin{figure}[t]  \centering
  \includegraphics[width=\textwidth]{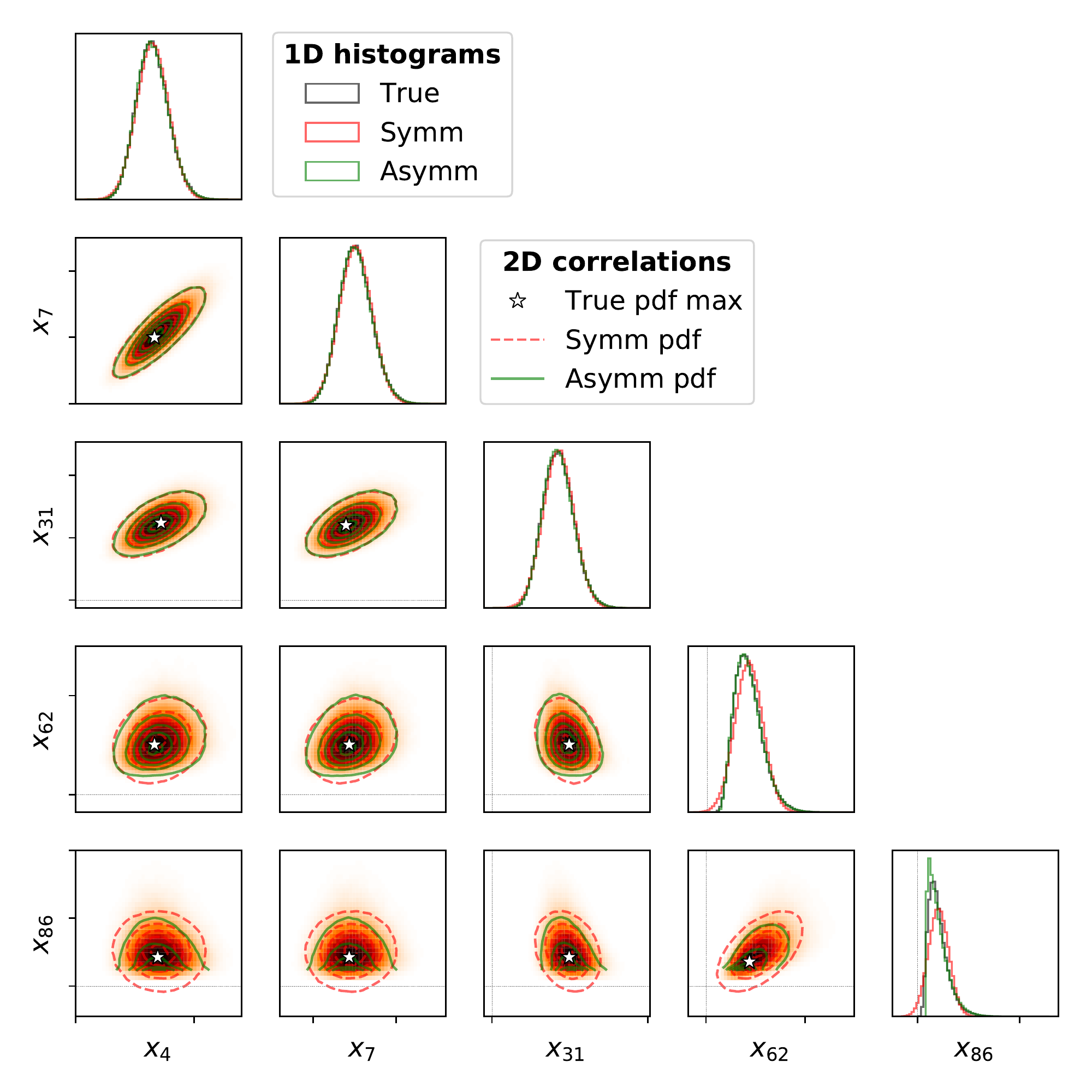}
  \caption{2D distributions of $\hat{n}_{b,I}$ against $\hat{n}_{b,J}$ for the LHC-like 
    experimental pseudo-search  
    as described in the text. The
    background heat map is generated from 100,000 samples from the true 
    model, the dashed red contours from the symmetric SL, and the solid
    green contours from the asymmetric SL. The diagonal panels show the
    1D distribution in each of the bins for the toys (black histograms), and the
    symmetric (red histograms) and asymmetric (green histograms) SLs.
    In the pair of
    high-statistics bins in the top-left plot, clear agreement is seen between
    the symmetric and asymmetric  SLs; in the top-right, deviations start to
    appear, and in the low-statistics bin~$J=86$ of the bottom plot the
    asymmetry is seen to become very significant, and the symmetric  SL form has a
    significant probability density fraction in the negative-yield region.}
  \label{fig:distributions2d}
\end{figure}

In Figure~\ref{fig:distributions2d}, 2D projections of the background
distributions are shown between four pairs of signal-region bins: bin pair
$(4,7)$ shows a projection for high-statistics bins where both the asymmetric and symmetric  SL agree closely with the true distribution (that obtained in the pseudo-datasets);
the true distribution in $(4,62)$ starts to display deviations from the multivariate
normal approximation which are well captured by the asymmetric SL. This is
expected when the skew, defined as $m_{3,I}/(m_{2,II})^{\frac{3}{2}}$, is small.
However, in the bottom pair of plots with bins~4 and~62 joint with the low-statistics
bin~86, the proximity of the mean rate to zero induces a highly asymmetric
Poisson distribution which neither SLs can model well. In these last
two plots, it can be seen that the asymmetric SL peaks at too low a value,
near a sudden cutoff also seen in Figure~\ref{fig:distributions}, while the
symmetric SL peaks at too high a value. 
In this region a better modelling would  require
evaluation of higher-order coefficients (and/or off-diagonal skew terms) and
hence higher moments of the experimental distributions.

An advantage of the asymmetric SL is that a strictly positive
approximate distribution can be guaranteed, while the symmetric SL can have a
significant negative yield fraction as seen in the figures for bin~86. Sampling
from the symmetric SL, \eg\ for likelihood marginalisation, requires that the
background rates be positive since they are propagated through the Poisson
distribution. The asymmetric SL provides a controlled solution to this issue,
as opposed to \emph{ad hoc} methods like use of a log-normal distribution or
setting negative-rate samples to zero or an infinitesimal value: the symmetric SL has a negative fraction of $\sim\!11.6\%$, while the
asymmetric SL has a negative fraction of exactly zero.

Typically in searches for new physics, limits on models for new physics
are determined using ratios of the likelihood at different values of the parameters of interest.
In the simplest case, a single parameter of interest is defined as $\mu$, often referred to
as the signal strength, which multiplies the expected contribution, under some specific signal hypothesis,
of the signal across all regions of the search, giving,

\begin{equation}
 n_{s,I}(\bm{\alpha}) = \mu n_{s,I},
\label{eq:muscale}
\end{equation}
where the yields $n_{s,I}$ here refer explicitly to the expected contributions from signal for a specified hypothesis.
In order to remove the dependence of the likelihood on the nuisance parameters, $\bm{\theta}$, the nuisance parameter values are set to those at which
the likelihood attains its maximum for a given set of $n^{\rm{obs}}_{I}$. This is 
commonly referred to as ``profiling'' over the nuisance parameters\footnote{Other 
procedures, such as marginalisation, can also be used to remove the dependence on 
the nuisance parameters. For 
reviews on how likelihoods, such as the simplified likelihood presented here, are
used in searches for new physics, see Refs.~\cite{CowanPDGProb,0954-3899-45-3-033001}}. 

\begin{figure}[t!]
  \centering
  \includegraphics[width=0.66\textwidth]{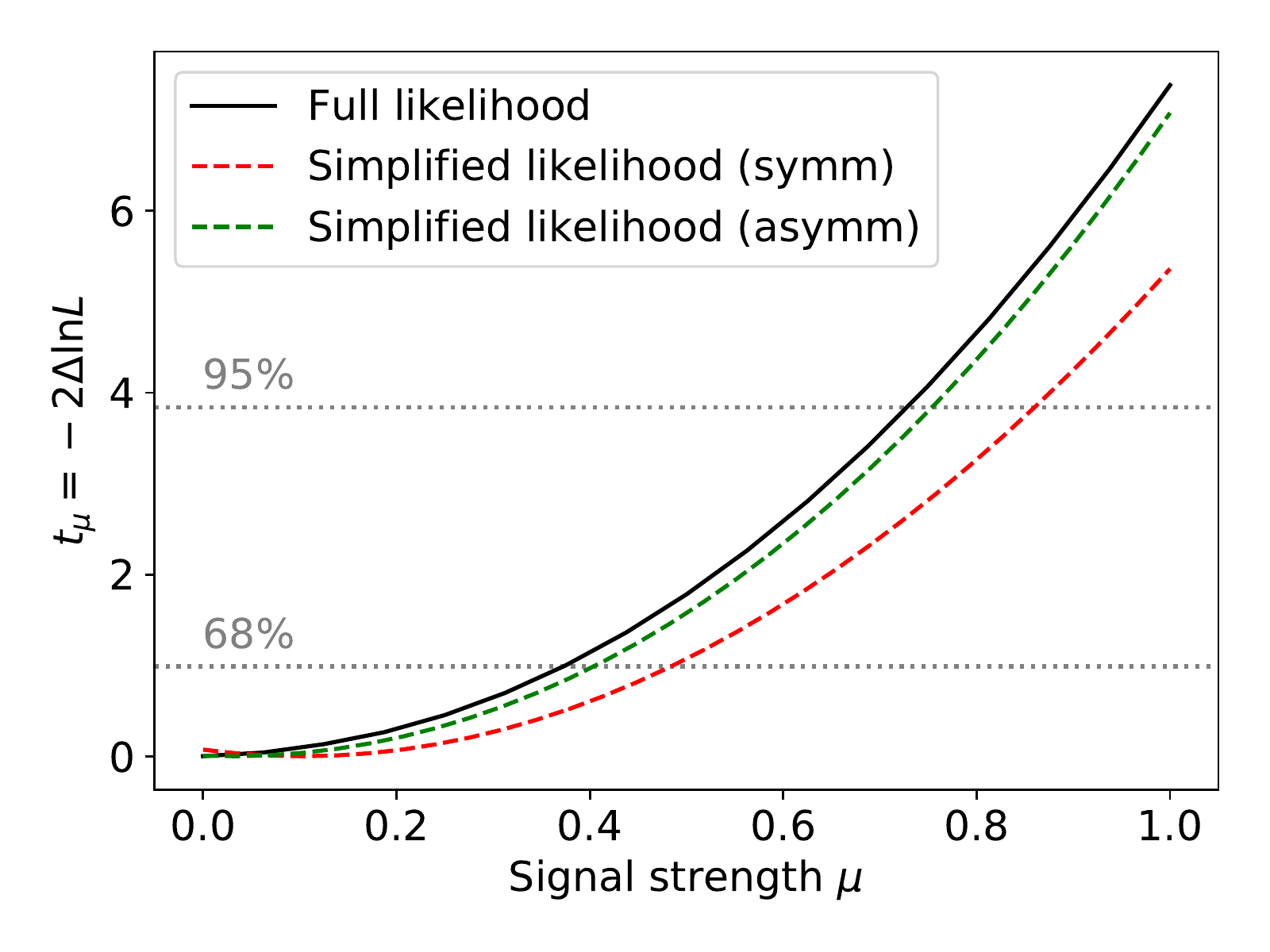}
  \caption{Value of $t_{\mu}$ as a function of $\mu$ for the pseudo-search assuming the experimental likelihood (black solid line) and simplified likelihood retaining (green dashed line) or not (red dashed line) the contribution from the quadratic term. The horizontal lines drawn at $t_{\mu}=1$ and $3.86$ represent the
  values for which the 68\% and 95\% CL exclusions can be determined, assuming certain asymptotic properties of the distribution of $t_{\mu}$.}
  \label{fig:tmucompare}
\end{figure}

\begin{equation}
L_{\rm{S}}^{\rm{max}}(\mu) = \rm{max}_{\theta_{I}}\left\{ L_{\rm{S}}(\mu,\bm{\theta}) \right\}.
\end{equation}
The test-statistic $t_{\mu}$ is then defined using the ratio,

\begin{equation}
t_{\mu} = -2\ln \frac{L_{\rm{S}}^{\rm{max}}(\mu)}{L_{\rm{S}}^{\rm{max}}},
\end{equation}
where $L_{\rm{S}}^{\rm{max}}$ denotes the maximum value of $L_{\rm{S}}^{\rm{max}}(\mu)$ for any value of $\mu$.\footnote{The precise definition of the test-statistic used
as searches at the LHC and the procedures used to determine limits are slightly different to that presented here and are detailed in Ref.~\cite{CMS-NOTE-2011-005}.}
Similarly, such likelihood ratios are also used for quantifying some excess in the case of the discovery
of new physics~\cite{CMS-NOTE-2011-005}.
The test-statistic can also be constructed for the experimental likelihood $L(\mu,\bm{\delta})\pi(\bm{\delta})$, where the same substitution as in Eq.~\eqref{eq:muscale} is applied,
by profiling the elementary
nuisance parameters $\bm{\delta}$. A direct comparison of the test-statistic for the full and simplified likelihoods, as a function of $\mu$, 
is therefore possible.

Figure~\ref{fig:tmucompare} shows a comparison of the value of $t_{\mu}$ as a function of $\mu$ for the pseudo-search between the full (experimental)
likelihood and the asymmetric SL. In addition, the result obtained using only the symmetric SL is shown. As expected, the
agreement between the full and simplified likelihood is greatly improved when including the quadratic term. A horizontal line is drawn at the value of
$t_{\mu}= 3.86$. The agreement in this region is particularly relevant due to the fact that asymptotic approximations for the distributions of $t_{\mu}$~\cite{Cowan:2010js} allow one to determine the 95\% confidence level (CL) upper limit on the signal strength, $\mu_{\rm{up}}$.
The signal hypothesis is ``excluded'' at 95\% CL if $\mu_{\rm{up}} < 1$.

When determining the SL coefficients, we have relied on pseudo-datasets, as we expect this will often be the case for anyone providing SL 
inputs for real analyses. The accuracy of the SL coefficients will necessarily depend on the number of pseudo-datasets used to calculate them. 
To investigate this, we have performed a study of the rate of convergence of the SL coefficients by calculating them using several different 
numbers of pseudo-datasets, the largest being 100,000 pseudo-datasets.  
The coefficients for the three bins calculated using 100,000 pseudo-datasets are; $a=84.9,~b=8.27,~c=0.32$ for bin 4, 
$a=2.61,~b=0.90,~c=0.11$ for bin 62, and $a=0.90,~b=0.47,~c=0.13$ for bin 86. 
The calculation of the coefficients is repeated using many independent sets of a fixed number of pseudo-datasets, resulting in a distribution of 
calculations for each coefficient. 

\begin{figure}[t!]
  \centering
  \includegraphics[width=0.34\textwidth]{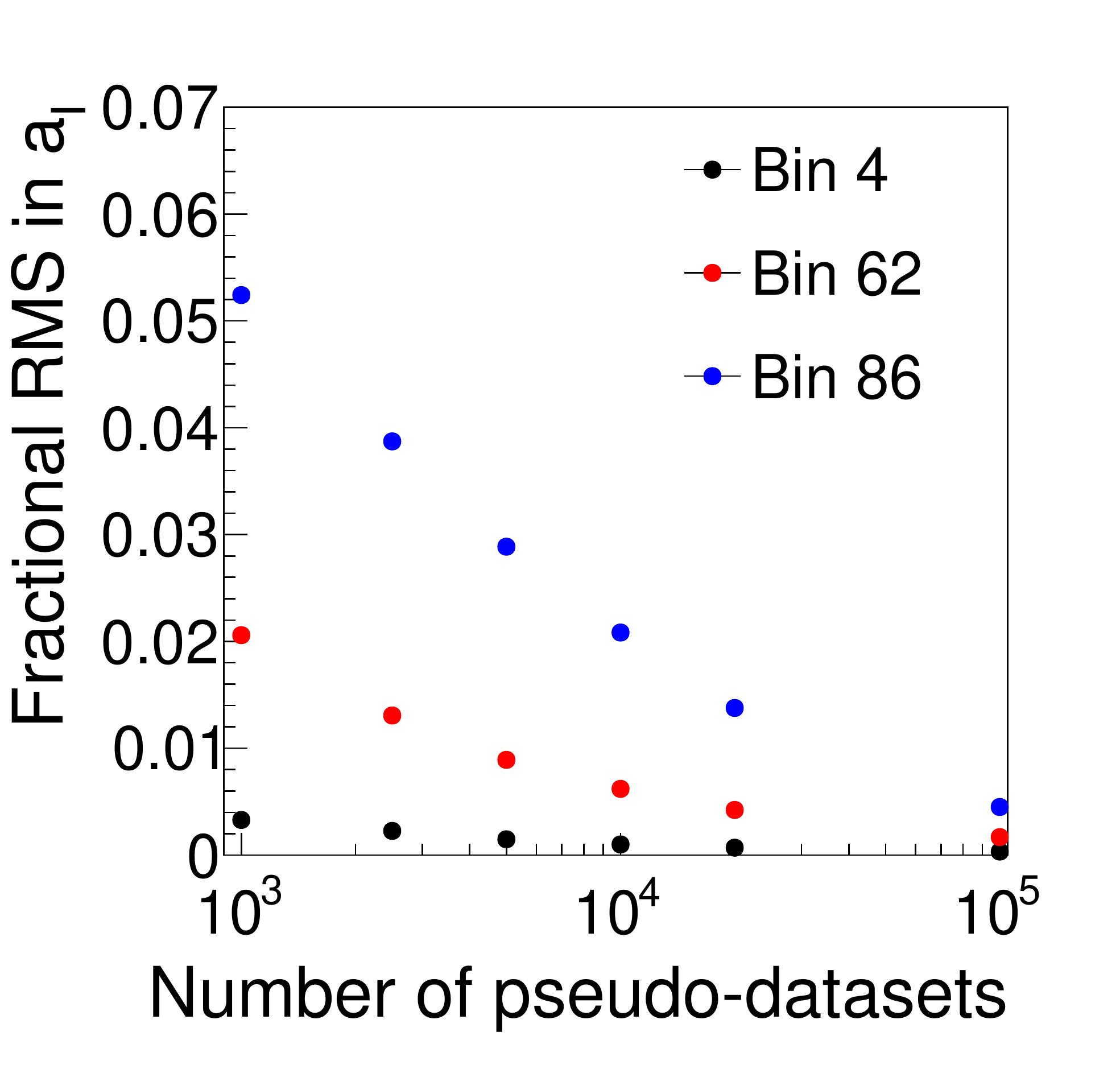}%
  \includegraphics[width=0.34\textwidth]{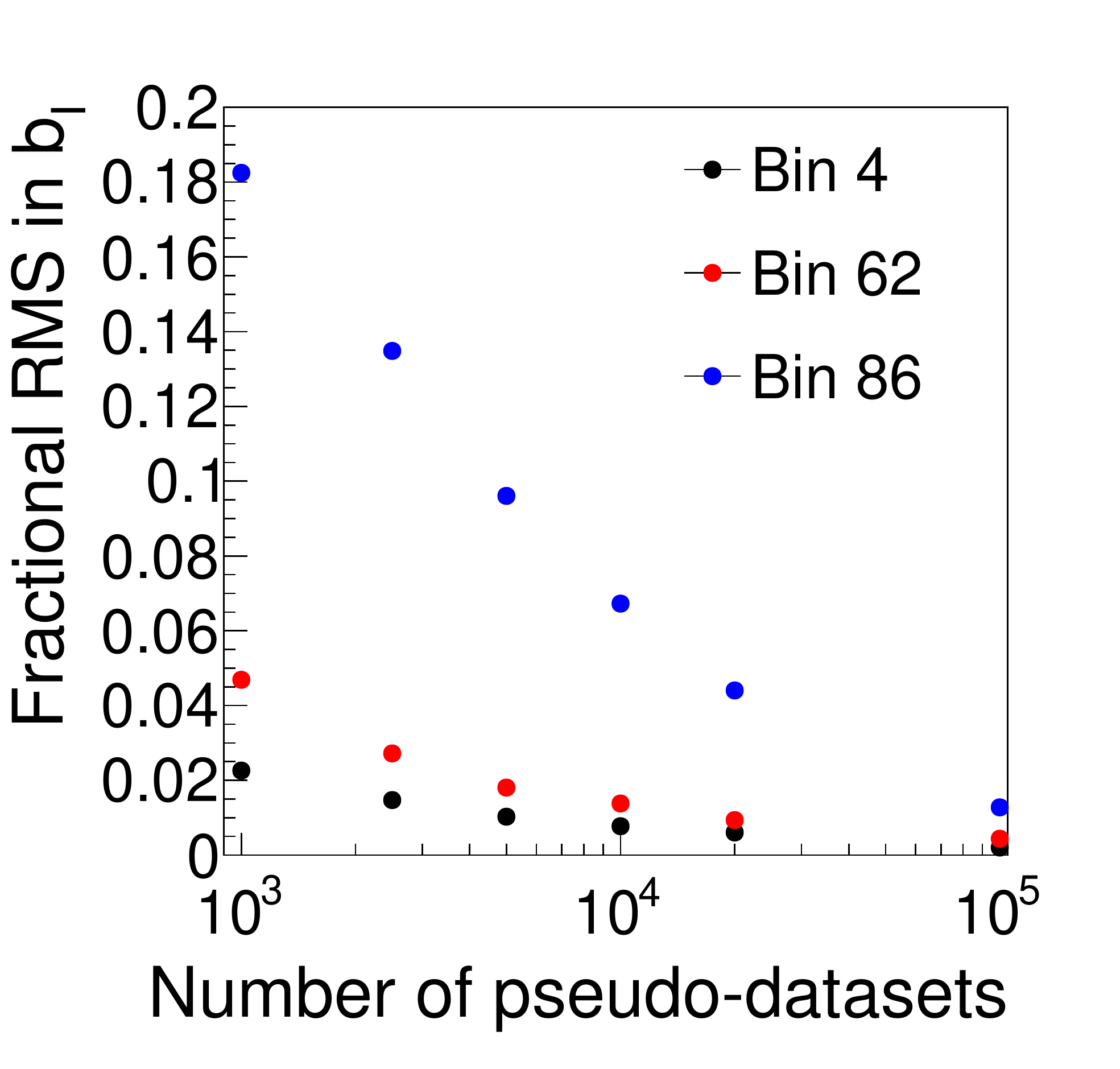}%
  \includegraphics[width=0.34\textwidth]{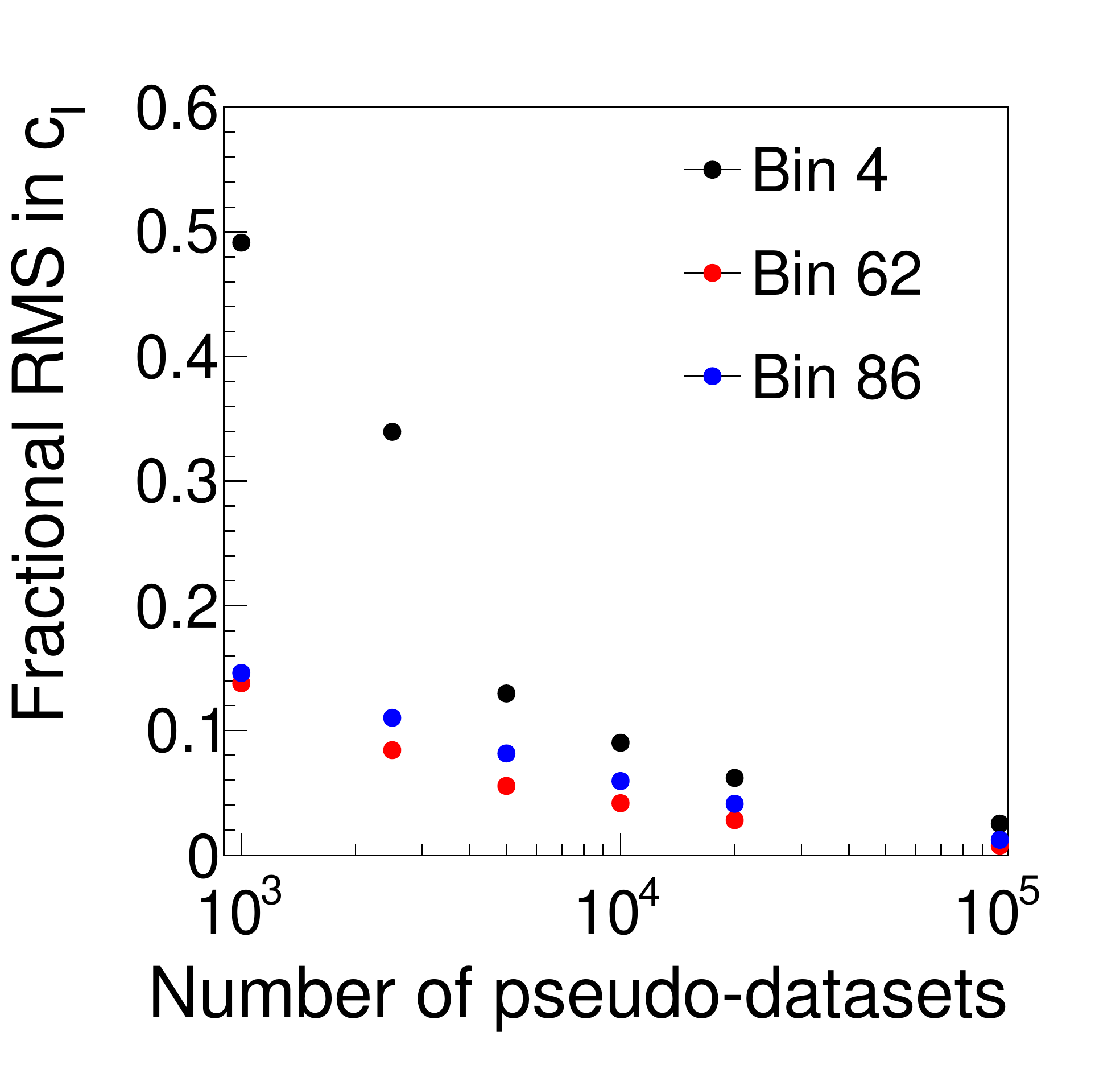}
  \caption{RMS of the SL coefficients relative to the mean
    coefficient value determined from 100,000 pseudo-datasets for $a_{I}$
    (left), $b_{I}$ (middle), and $c_{I}$ (right). The distributions are shown
    for $I=4$ (black points), $I=62$ (red points) and $I=86$ (blue points).}
  \label{fig:SLConvergence}
\end{figure}

The root mean square (RMS)  of the resulting distributions provides an estimate for how much variation can be expected in the calculation 
of the SL coefficients given a limited pseudo-data sample size. 
The RMS values are normalised to the RMS of the distributions resulting from a sample size of 
100,000 pseudo-datasets to give a relative RMS. 
The relative RMS of the distribution of the 
coefficients calculated using increasing numbers of pseudo-datasets is shown Figure~\ref{fig:SLConvergence}.

The coefficients $a$ and $b$ can be calculated with relatively high precision using only 1000 pseudo-datasets in each case. This is true 
whether the value of $b$ is large compared to $a$, as in the case of bin 86, or not, as in the case of bin 4. The determination of 
the $c$ coefficient for bin 4 
however is slower to converge, requiring 5000--10,000 pseudo-datasets to calculate accurately. However, since the value of $c$ for this bin is 
relatively small compared to $b$, the coefficient $c$ is less relevant so that a poor accuracy will have little effect on the accuracy of 
the SL. In bin 86, the value of $c$ is relatively large, compared to $b$, meaning it will significantly contribute to the SL. In this case, 
the convergence is quite fast, with only 2,500 pseudo-datasets required to achieve a 10\% accuracy in the value of $c$. We find the property 
that bins with large $c$ values, compared to $b$ values, require fewer pseudo-datasets to achieve a good accuracy than bins for which the $c$ value 
is less relevant generally holds in this study.

\section{Summary and conclusions}
\label{se:conclusions}

The transmission of highly complex LHC likelihoods from the experimental collaborations to the scientific community
has been a long standing issue. 
In this paper, we proposed a simplified likelihood framework which can account for non-Gaussianities as a convenient way of presentation 
with a sound theoretical basis. 

Although the SL is  accurate, it is still an approximation of the full experimental likelihood, hence the collaborations 
do not have to release their full 
model.
Meanwhile, for the public, having a good approximation of the true likelihood is sufficient for most phenomenology purposes. 
Moreover, the SL is very simple to transmit, requiring neither a substantial  effort for the experimentalists to release it nor for the user to construct it. Additionally, with some standardisation effort, part of this transmission process can be automated. 

In this paper we introduced the formalism for the asymmetric version of the SL. 
This formalism follows directly from the central limit behaviour of the combination of systematic uncertainties: asymmetry is recognised as the subleading term of the asymptotic distribution dictated by the CLT, which is then recast in a convenient form in the SL formulation. 
The inclusion of asymmetry completes the SL and provides a fully reliable framework.

The asymmetric SL can be built either from the elementary systematic uncertainties themselves or from the three first moments  of the combination of the systematic uncertainties, which are easily obtained via MC generators. 
Using a realistic LHC-like pseudo-search for new physics, we demonstrated that including asymmetry in the SL provides  
an important gain in accuracy, and  that it is unlikely that higher moments will be needed.

The SL formalism discussed in this paper focusses on datasets with more systematic uncertainties than observables
 (\textit{i.e.} $N\geq P$), and a few extra simplifying approximations have been made.  
The conditions of its validity are summarised as follows:
\begin{itemize}
\item \textit{Convergence of the central limit theorem}: 
There should be enough independent sources of uncertainties for the combined distribution to tend towards a Gaussian. 
This is the fundamental condition underlying the SL approach. The leading, asymmetric corrections to the Gaussian 
can be treated as described in this work.
\item \textit{Sufficiently symmetric combined uncertainties}: Alltough we have consistently included skewness in our formalism, it cannot be arbitrarily large, as discussed in Section~\ref{se:precision}. In particular the formulas used to derive the SL coefficients are valid only when the second ($m_{2,II}$) and third ($m_{3,I}$) moments satisfy $8m^{3}_{2,II}\geq m^{2}_{3,I}$. 
\item \textit{Negligible signal uncertainties}: In order to be re-usable for different signal hypotheses, e.g., for limit setting on different models, 
the SL must not depend on the parameters of interest. This is ensured if signal uncertainties are negligible to good approximation.    
While the inclusion of pure signal uncertainties is straightforward, systematic uncertainties that are correlated between signal and background must be included in the derivation of the SL coefficients. Without this, the simplification to Eq.~\eqref{eq:SL_LHC} is no longer valid, as discussed in Section~\ref{se:application}. 
\end{itemize}

In practice, for the transmission of the SL data from an experiment to the public, our recommendation is to simply release the three first moments of the combined uncertainties, preferably via the HepData repository in the error source format.  The SL framework is flexible in the sense that it can apply to one or more subsets of the systematic uncertainties,  and the HepData error source format  has adequate flexibility to account for any partitions of the uncertainties the releaser wishes to make.

If adopted by the experimental and theory communities, and provided the above validity conditions are respected, 
the SL framework has the potential to considerably improve both the documentation and the re-interpretation of the LHC results.

\section*{Acknowledgements}

This work has been initiated at the \textit{LHC Chapter II: The Run for New Physics} workshop held at IIP Natal, Brazil,  6--17 Nov.\ 2017. 
We thank the IIP Natal for hosting the workshop and creating a most inspiring working atmosphere.

AB is supported by a Royal Society University Research Fellowship grant. 
MC is supported by the US Department of Energy under award number DE‐SC0011702.
SF is supported by the S\~ao Paulo Research Foundation (FAPESP) under grants \#2011/11973, \#2014/21477-2 and  	
\#2018/11721-4. 
SK is supported by the IN2P3 project ``Th\'eorie LHCiTools'' and the CNRS-FAPESP collaboration grant PRC275431. 
NW is funded through a Science and Techologies Facility Council (STFC) Fellowship grant \#ST/N003985/1.

\appendix

\section{The CLT at next-to-leading order}
\label{app:skew}

Let us show in a 1D example how the skew appears  in the asymptotic distribution. Consider $N$ independent centered nuisance parameters $\delta_j$ of variance $\sigma^2$ and third moment $\gamma$. Define \be Z=\frac{\sum_{j=1}^N \delta_j}{\sqrt{N}}\,.
\ee
The characteristic function of $Z$ is given by
\be
\varphi_Z(t)=\prod_{j=1}^N\varphi_{j}\left(\frac{t}{\sqrt{N}}\right),
\ee
where $\varphi_{j}(x) = {\bf E}[e^{ix\delta_{j}}]$.
 In the large $N$ limit, each individual characteristic function has the expansion
\be
\varphi_{j}\left(\frac{t}{\sqrt{N}}\right)= 1-\frac{\sigma^2 t^2}{2N}-i \frac{\gamma t^3}{6 N^{3/2}} +O\left(\frac{t^4}{N^2}\right)\,.
\ee
It follows that the full characteristic function $\varphi_Z$ then simplifies to
\be
\varphi_Z(t)=\exp\left(-\frac{\sigma^2 t^2}{2}-i \frac{\gamma t^3}{6 \sqrt{N}} +O\left(\frac{t^4}{N}\right)\right) \label{eq:CF_CLT}
 \ee
 This characteristic function is simple but has no exact inverse Fourier transform.

To go further, let us observe that the $Z$ random variable could in principle be written in terms of a normally distributed variable $\theta\sim {\cal N}(0,\sigma^2)$,
 with $Z=\phi(\theta)$ where $\phi$ is a mapping which is in general unknown.  At large $N$ however, we know that $Z$ tends to a normal distribution hence $\phi$ tends to the identity. Thus we can write $Z=\sqrt{N}\phi\left(\frac{\theta}{\sqrt{N}}\right)$ and Taylor expand for large $N$,
\be
Z=\theta+\frac{c}{2\sqrt{N}}\theta^2+O\left(\frac{1}{N}\right)\,.
\ee
Let us now compare the characteristic function of  this expansion to Eq.~\eqref{eq:CF_CLT}.
We find that the characteristic function is given by
\be
\varphi_Z(t)={\bf E} \left[ \mathrm{e}^{it \left(\theta+\frac{c}{2\sqrt{N}}\theta^2+O\left(\frac{1}{N}\right)\right)} \right]
=\exp\left(-\frac{\sigma^2 t^2}{2}-i \frac{c t^3}{ 2 \sqrt{N}} +O\left(\frac{1}{N}\right)\right)
\label{eq:CF_exp}
\ee
after using the large $N$ expansion. This function matches Eq.~\eqref{eq:CF_CLT} for $c=\frac{\gamma}{3}$. Thus we have found the normal expansion provides a way to encode skewness in the large $N$ limit. Namely, we find that the $Z$ variable converges following
\be
Z\rightarrow \theta+\frac{\gamma}{3\sqrt{N}}\theta^2\,,\,\,N\rightarrow \infty  \quad \textrm{with } \quad \theta\sim{\cal N}(0,\sigma^2)\,.
\ee
When the quadratic term becomes negligible the distribution becomes symmetric, and we recover the usual CLT.
We can see that  for finite $N$ (as opposed to $N\rightarrow \infty$) the support of $Z$ is not $\bf R$. For example for $\gamma>0$, we have
$Z > -3\sqrt{N}/4\gamma$.

\section{Reference Code}
\label{app:reference_code}

A reference implementation in Python code, {\tt simplike.py}, is provided in
\begin{quote}
  \url{https://gitlab.cern.ch/SimplifiedLikelihood/SLtools}. 
\end{quote}
It includes functions to
calculate the SL $a_I$, $b_I$, $c_I$, and $\rho_{IJ}$ coefficients from provided
moments $m_{1,I}$, $m_{2,IJ}$ and $m_{3,I}$; and an \texttt{SLParams} class
which computes these and higher-level statistics such as profile likelihoods,
log likelihood-ratios, and related limit-setting measures computed using
observed and expected signal yields. For convergence efficiency, the profile
likelihood computation makes use of the gradients of the SL log-likelihood with
respect to the signal strength $\mu$ and nuisance parameters $\bm{\theta}$,
which we reproduce here to assist independent implementations:
\begin{align}
  \begin{split}
    \ln\! \big( L_{\rm{S}}(\mu,\bm{\theta} )\pi(\bm{\theta}) \big) =&
    \sum_I^P \Big[ n^{\rm{obs}}_{I} \ln \left(\mu n_{s,I} + n_{b,I}({\bm \theta}) \right) - \left( \mu n_{s,I} + n_{b,I}({\bm \theta}) \right) - n^{\rm{obs}}_{I}! \Big]\\
    &- \frac{1}{2} \bm{\theta}^\mathrm{T} \bm{\rho}^{-1} \bm{\theta} - \frac{P}{2} \ln 2\pi
  \end{split}\\[2ex]
  \frac{\partial\ln L_{\rm{S}}}{\partial\mu} =& \sum_I^P \left( \frac{n^{\rm{obs}}_I}{\mu n_{s,I} + n_{b,I}(\bm{\theta})} - 1 \right) \cdot n_{s,I} \\
  \frac{\partial\ln L_{\rm{S}}}{\partial\theta_{\!A}} =& \left( \frac{n^{\rm{obs}}_A}{\mu n_{s,A} + n_{b,A}({\bm \theta}) } - 1 \right) \cdot \big( b_A + 2 c_A \theta_{\!A} \big) - \sum_I^P \rho_{\mspace{-1mu}AI}^{-1} \, \theta_I ~,
\label{eq:SL_LHC_refcode}
\end{align}
where $n_{b,I}({\bm\theta}) = a_{I} + b_{I}\theta_{I} + c_{I}\theta_{I}^{2}$.

The reference code has been written with reverse engineering and
comprehensibility of the calculations explicitly in mind. While it computes
likelihood statistics on a reasonable timescale, further (but less readable)
optimisations can be added for production code.

A demo of the construction of the simplified likelihood, and profiling as a function of a signal strength parameter, is given in {\tt simplikedemo.py}.
Finally, the SL pseudo-data are available on the HepData repository at \url{https://www.hepdata.net/record/sandbox/1535641814}.

\bibliographystyle{jhep}
\bibliography{biblio}

\end{document}